# A photon model based upon chaos produced by static, Schwinger-level electric field nonlinearities that satisfies all first-order properties

Dale M. Grimes[1] and Craig A. Grimes[2]
[1] The Pennsylvania State University, University Park PA 16802 United States. Email: dale.grimes@gmail.com
[2] Raleigh, North Carolina. Email: craig.grimes40@gmail.com

## Abstract

In this work we postulate that Schwinger's threshold for a dynamic electric field intensity to induce spatial nonlinearity is a special case and, more generally, it is the threshold field for both static and dynamic electric fields. Fields of this magnitude induce negative energy charges to adapt positive energy attributes; within an atom they also support inter-state energy transfers and intra-state chaotic mixing of time-varying fields. Nonlinearity-induced chaos forms the basis for the probabilistic nature of photon creation. Answers to physical problems at atomic and lower scales continuously evolve because chaotic-like electron movements change their configurations on a time scale of 10 zs. Within atoms, frequency mixing that creates an optical frequency field occurs in the nonlinear region surrounding the nucleus. On a probabilistic basis a ring of vacuum charge can be induced that forms into an equivalent waveguide that confines the energy as it travels permanently away from the atom. The propagating relativistically augmented fields losslessly induce charges that bind and protect the energy carrying fields. The photon charge-field ensemble, which we show is localizable, is thermodynamically closed and possesses all first-order photon properties including zero rest mass and permanent stability. For near neighbor photons travelling at a speed approaching but not reaching $c$ we find a small, constant, attractive force between photons with circularly antiparallel polarization.

## 1. INTRODUCTION

Photon properties are of ongoing scientific interest [1-14], with commercial applications that include optical communication, temporal imaging, and supercontinuum generation. It has been noted [15] that the amount of information carried by a photon is potentially enormous, and utilizing this information would enable quantum communication systems with extraordinary capacities and exceptional levels of security [16,17]. This work presents a unique photon model that details structure, propagation, and spontaneous generation. It is an interdisciplinary study of atomic and optical phenomena based upon techniques selected from physics, electrical engineering, and optics. We present our ideas, conclusions, and the thoughts that led us to them with the hope it will assist others with their research and development.

Schwinger calculated that a dynamic electromagnetic field of $1.3 \times 10^{18}$ V/m is the threshold between which the vacuum of space presents low-field linear and high-field nonlinear responses [18]. To foster a more complete understanding of this effect many capable experimentalists have attempted to create a Schwinger threshold field in the laboratory using lasers [19-29], but have been unsuccessful. They explain the difficulty as nonlinearity-induced charge transitions and resulting unavoidable effects that extract energy. This leaves characteristics of high intensity fields as a largely unexplored regime.

Moderate intensity physical phenomena such as waves and particle interactions at a distance are linear phenomena, and well understood on the basis of superposition with other linear phenomena, but the superposition principle cannot be used to construct particles. For example, nonlinearities are essential for the creation of a lepton pair from a high-energy photon. The desire to understand particle formation has resulted in many and extensive studies to determine results of adding nonlinear terms to known linear equations, but without widespread success. Although Schwinger's threshold field, as derived, applies only to dynamic electromagnetic fields we *postulate* that it is a special case of the general rule that threshold-level electromagnetic fields, static and dynamic, force a nonlinear response from the spatial vacuum that induces negative-energy charges toward positive energies and thus prevent any field, including static ones, from exceeding his calculated threshold value. We then apply conventional linear physics to determine results, and report them in this paper.

Jackson [30] pointed out that static nuclear fields near atomic nuclei are as large as $10^{21}$ V/m. Indeed, regions with a calculated Coulomb field exceeding the threshold field are ubiquitous and centered on every atomic nucleus. In many cases the nonlinear region extends to distances of 150 fm and Schrödinger's equation shows that a portion of every atomic eigenstate lies within that region. Since atomic emission satisfies the Manley-Rowe equations [31] and since they are trivial in linear media, we take the agreement as evidence of nonlinearity in atoms. Since each eigenstate contains a nonlinear volume and since eigenstate electrons are dynamic entities, we anticipate chaotic-like behavior of electrons subject only to the constraints of atomic conservation laws.

Combining the idea of charge induction from vacuum with the theory of waveguides provides a means of creating an ordered, hybrid charge - electromagnetic field structure that exhibits first-order properties of photons. Since photon construction requires a confluence of events its completion is probabilistic, but with satisfactory conditions a ring of charge proportional to cos$\phi$, centered on the axis of propagation, is induced that supports and guides the electromagnetic fields. With propagation the ring of charge extends, with the leading edge of the fields, becoming a circular cylinder that supports a thermodynamically stable energy packet. Only the photon energy propagates; charges are induced in position by the fields, retained in position as they bind and guide the passing fields, and losslessly return to negative energy states after photon passage. The calculated photon size is sufficiently flexible to include frequency information coded either by energy or wavelength.

Herein, the model we present shows that both classical electromagnetism and a disruption of the local three-dimensional spatial continuum are essential for a photon's existence. Our results show the photon is intrinsically both a wave and a particle [32-43], wherein the 'particle' is charge induced by the fields of the photon, which exceed the Schwinger nonlinearity threshold, from the Dirac vacuum. In agreement with Ojima and Saigo [10] we find photons comprise a composite system with coupled dynamics.

In **Section 2** we discuss the chaotic nature of the fields and charges within an atom, in **Section 3** we discuss a thermodynamically-closed charge-field ensemble with known first order photon properties, and in **Section 4** discuss spontaneous emission, induced emission, and photon size. In **Section 5** we examine the force between two near-neighbor photons propagating in the same direction identical in all respects except in one case the spins have the same direction and in the second case the spins have opposite directions. For each case we find that there is a small force between them: the force is time-varying with parallel spins, and attractive with antiparallel spins.

## 2. FIELDS AND CHARGES OF AN ATOM

With $c$ representing the speed of light, $\hbar$ the reduced Planck's constant, and $m$ and $e$ electron mass and charge, the threshold field is [18]:

$$E_S = m^2 c^3 / e\hbar \simeq 1.3 \times 10^{18} \, \text{Vm}^{-1} \tag{1}$$

Several laser groups report attempts to create Schwinger's threshold field in the laboratory, but without success. The difficulty is that, in accordance with Dirac's theory, the fields force electron-positron pairs to tunnel to positive energies, separates them into independent existences, and then accelerates them in opposite directions. These actions extract the transformation energy and leave the science of extreme electric fields relatively unexplored [19-29]. In the absence of experimental information, we postulate that both static and dynamic fields force charge transformation from negative to positive energies, but static fields lack the dynamic quality required for lepton formation. With static fields, instead of a dynamic reaction the opposite surface charge density is retained in equilibrium positions between the repulsive creating field and attractive fields of their own making. Dittrich and Gies point to an equivalence between properties of dielectric media and quantum vacuum properties, with the Lamb shift and the Casimir effect as evidence [44,45]. Virtual pairs in vacuum and actual pairs in polarized media respond to applied fields similarly and, in some cases, the virtual pairs may include milli-charges [46-49]. Although event details near the threshold field intensity have not been explored, naturally occurring static fields of this magnitude occur in the immediate vicinity of all atomic nuclei [30]. An internal atomic nonlinearity creates power-frequency transitions in accordance with the Manley Rowe relationships [31]; we take this behavior as evidence that Schwinger's nonlinearity extends to static nuclear fields.

**Table I** lists relevant nuclear parameters and associated fields for six selected elements. The second column is atomic number $Z$, the third column is nuclear radius $R_N$, calculated using the formula $R_N = 1.07A^{1/3}$ where $A$ is the number of contained nucleons, the fourth column shows the calculated nuclear surface-to-threshold field ratio, with $E_N$ denoting the nuclear field at the surface of the nucleus, and the fifth column shows radius $R_S$ at which the field, calculated using nuclear charge and Coulomb's law, drops to the threshold value.

**Table I**: Relevant nuclear parameters for bare atomic nuclei.

|     | $Z$ | $R_N$ (fm) | $E_N/E_S$ | $R_S$ (fm) |
|-----|-----|-----------|-----------|-----------|
| Mg  | 12  | 3.10      | 138       | 115       |
| Ca  | 20  | 5.16      | 161       | 149       |
| Fe  | 26  | 4.09      | 169       | 170       |
| Rb  | 37  | 4.71      | 185       | 202       |
| Ba  | 56  | 5.52      | 208       | 249       |
| Hg  | 80  | 6.26      | 223       | 298       |

For atoms with a full complement of electrons and, with $r$ representing the radius, the actual value of $R_S$ shown is less than shown in **Table I** because of time-average electron charge at $r < R_S$; accuracy may be increased by including electron charge within the Schwinger region, $R_N < r < R_S$, particularly as calculated using descriptive Schrödinger wave functions as corrected [50,51].

From the perspective of classical physics Schrödinger's equation is based upon a Fourier integral transform between spatial and momentum spaces, and such transforms are valid if and only if both spaces are linear and at least piece-wise continuous. Therefore, the equation is and is not valid, respectively, within linear and nonlinear spaces, and the behaviors of charges and fields in the nonlinear medium are both unknown and unknowable. Within the nonlinear region we know only that the static field intensity equals $E_S$ and charge density exists throughout the region, but have no knowledge of the detailed ebb and flow of charge under the influence of added fields. Atomic radii are on the order of 100 pm and $R_S$ sizes are on the order of 200 fm. Although the nonlinearity occupies about $10^{-8}$ of an atom's interior space, we suggest it has a major influence on atomic interactions since only it is subject to chaotic mixing [52-54]. In the linear region charge is distributed throughout the eigenstates in a manner that conserves time and space averages of energy, linear momentum, and angular momentum as the charge distributions cycle through all possible formations. The intrinsic electron frequency and wavelength are $v_0 = mc^2/h = 7.8 \times 10^{20}$ Hz, $\lambda = h/mc = 386$ fm, hence charge configurations are perturbed at the rate of about $7.8 \times 10^{20}$ per second. An electron has no known components or substructure; we take the simplified view of an electron as an adaptive charged cloud that maintains said parameters. When free of constraints the electron becomes a sphere of charge, and within an eigenstate it expands to occupy the entire state in accordance with the wave function. As illustrated in Supplemental Information SI-2, interactions within the nonlinear region, in a small but continuous way, supply sufficient chaotic energy into each eigenstate to keep non-conserved atomic properties aperiodic.

The Manley Rowe power-frequency relationships govern the rate at which intra-atomic nonlinearities create inter-frequency energy exchanges [31]. We briefly remind the reader, since the relationships are critical to what follows, that during an interaction between the electrons of two eigenstates, with different energies and frequencies, either up or down conversion may result, but only with a concurrent energy transfer at their difference frequency. For example, with $P_1$ and $\omega_1$ and $P_2$ and $\omega_2$ representing respectively initial and final eigenstate powers and frequencies and $P$ and $\omega$ representing the generated radiation, the Manley Rowe equations are:

$$\frac{P_1}{\omega_1} + \frac{P_2}{\omega_2} = 0 \ \text{ and } \ \frac{P_1}{\omega_1} - \frac{P}{\omega} = 0 \tag{2}$$

These equations govern lossless oscillating systems; by convention power emission is positive. A time integral shows the energy-to-frequency ratio is constant between interacting systems.

## 2.1 The nonlinear region

By the postulated extension of Schwinger's results to static fields, threshold field $E_S$ for the onset of spatial nonlinearity applies to the static fields created by atomic nuclei. Retaining the field value at $E_S$ can only be accomplished by a negative charge layer about the nucleus and a positive charge density distributed throughout the full non-linear region. By the divergence theorem and with $\kappa$ representing charge density, $\wedge$ denoting a unit vector, $\varepsilon_0$ the permittivity of free space, and $r$ the radial distance from the center of the nucleus, the charge density is $\kappa = \nabla \cdot \left( \varepsilon_0 E_S \hat{r} \right) = 2\varepsilon_0 E_S / r$. The resulting charge density and total induced positive charge are:

$$\begin{aligned} \kappa &= 2\varepsilon_0 E_S / r \\ q_0 &= Ze\left(1 - R_N^2 / R_S^2\right) \end{aligned} \tag{3}$$

Since ± charges are induced in equal measure, the magnitude of the negative charge adjacent to the nucleus is $-q_0$. An additional field applied to the region would not affect the field magnitude, but it would affect $\kappa(r)$; as we shall show this idea appears key to radiation emission by atoms.

## 3. A ONE-DIMENSIONAL PULSE

An antenna cannot create a field with a large wavelength-to-size ratio that propagates outward in less than three dimensions, yet photons certainly lie within that wavelength-to-size ratio and propagate in one-dimension. Microwave techniques use waveguides to obtain one-dimensional propagation, yet empty space contains no obvious means to construct one. We construct a working model of a photon that is based upon and consistent with classical electromagnetism, supports one-dimensional waves, and accurately describes first-order properties of optical photons. Both experimental and quantum theoretical studies have investigated possible photon structures [55-62]. Quite differently from them, we utilize steady-state solutions of the electromagnetic equations to examine how one-dimensional flows of microwave fields are created and controlled, and then seek to determine if an atom could use a similar but scaled technique to generate and control optical radiation. In this section we detail the fields and associated layers of induced charge on the surface of a one-dimensional, circularly cylindrical dielectric waveguide of radius $b$.

Only for TEM modes does the speed of propagation approach $c$; all other propagation modes are significantly slower. Another characteristic of all except TEM waveguide modes is a longest possible propagating wavelength. Looking ahead to the results shown in **Section 4.3**, our calculated photon radius is so small, at frequencies of interest, that neither TE or TM modes will propagate. Therefore, we consider only TEM modes.

Electric potential $\Phi$ satisfies the wave equation $\nabla^2 \Phi - \partial^2 \Phi / \partial t^2 = 0$, with $k$ the separation constant between space and time solutions, propagation in the $z$-direction obeys $\partial^2 \Phi / \partial z^2 + k^2 \Phi = 0$, the time and space dependence is $e^{i(\omega t - kz)}$ where $\omega$ is the radian frequency. The potential in the transverse plane satisfies the Laplacian equation, to which, using cylindrical coordinates $(\rho, \phi, z)$ the solutions are potentials $\Phi = \rho e^{-j\phi}$ and $e^{-j\phi}/\rho$; we are concerned with both. Both $i$ and $j = \pm(-1)^{1/2}$, with $i$ associated with space-time and $j$ the azimuth angle.

With bold face indicating vector the full set of fields, both internal and external, guided by a thin dielectric tube of radius $b$ is:

$$\boldsymbol{E}_{in} = \boldsymbol{E}_0 \left( \hat{\rho} - j\hat{\phi} \right) e^{-j\phi} e^{i(\omega t - kz)} \quad : c\boldsymbol{B}_{in} = j\boldsymbol{E}_{in}$$

$$\boldsymbol{E}_{ex} = -\boldsymbol{E}_0 \frac{b^2}{\rho^2} \left( \hat{\rho} + j\hat{\phi} \right) e^{-j\phi} e^{i(\omega t - kz)} : c\boldsymbol{B}_{ex} = -j\boldsymbol{E}_{ex}$$

$$(4)$$

With (4), making the equality $j = 0$ or $j = \pm i$ yields, respectively, linearly or circularly polarized fields. The flux lines for $\rho < b$ are straight lines and for $\rho > b$ form a circular arc.

With both interior and exterior fields present the charge density $\kappa$, current density $I$, and fields within the interface at radius $b$ are:

$$\boldsymbol{E}_b = -jE_0 \hat{\phi} e^{-j\phi} e^{i(\omega t - kz)} \qquad c\boldsymbol{B}_b = jE_0 \hat{\rho} e^{-j\phi} e^{i(\omega t - kz)}$$

$$\kappa = 2\varepsilon_0 E_0 \, e^{-j\phi} e^{i(\omega t - kz)} \qquad \eta \boldsymbol{I} = 2E_0 \hat{z} e^{-j\phi} e^{i(\omega t - kz)}$$

$$(5)$$

**Figure 1** depicts both inner and outer field forms and interfacial charge density $\kappa$. The charge and current densities are essential for the system to function; its value is determined by the difference in electric field intensities across the interface as the wave propagates, although individual charges remain *in situ* the magnitudes and phases propagate and satisfy (5). In the reference frame of the waveguide, although the actual $z$-directed charge motion is zero, current (5) is created by wave propagation past charge density $\kappa$ at speed $c$.

With circular polarization the Poynting vector $N$, field energy $W$, and linear momentum $M_z$, supported by length $l$ of the tube are:

$$\boldsymbol{N}_{in} = \hat{z} E_0^2 / \eta \quad : \quad W = \pi \varepsilon E_0^2 b^2 l \quad : \quad M_z = W/c$$

$$(6)$$

$$\boldsymbol{N}_{ex} = \hat{z} \frac{E_0^2}{\eta} \frac{b^4}{\rho^4} \quad : \quad W = \pi \varepsilon E_0^2 b^2 l \quad : \quad M_z = W/c$$

With the gauge in which the fields are functions of only the vector potential, $A_\phi = E_\phi / i\omega$ where $E_\phi$ is that of (5). With asterisk representing complex conjugate, the linear momentum per unit length equals product $A\kappa/2$. The angular momentum $\boldsymbol{L}$ about the $z$-axis is equal to the line integral of radius $b$ times $A\kappa/2$:

$$\boldsymbol{L} = -2\hat{z}\frac{ij}{\omega}\pi\varepsilon_0 E_0^2 b^2 l = -ij\frac{W}{\omega}\hat{z} \tag{7}$$

Retaining only the real part of (7) with respect to $j$ or writing $j = \pm i$ the energy-to-angular momentum ratios are respectively $W/L = \infty$ or $W/L = \pm \omega$. The linear and angular momenta result, respectively, from the field-flux product and the field-charge product.

The full interfacial boundary conditions are equal magnitude antiparallel field components $E_\rho$ and $B_\phi$, and equal magnitude parallel field components $E_\phi$ and $B_\rho$. The fields of (4) satisfy these conditions and the uniqueness theorem assures that the coaxial field ensemble described above *uniquely* possesses these properties: no other field forms can meet the same constraints. The external fields extend outward and mix with the environment but remain attached to the originating charges.

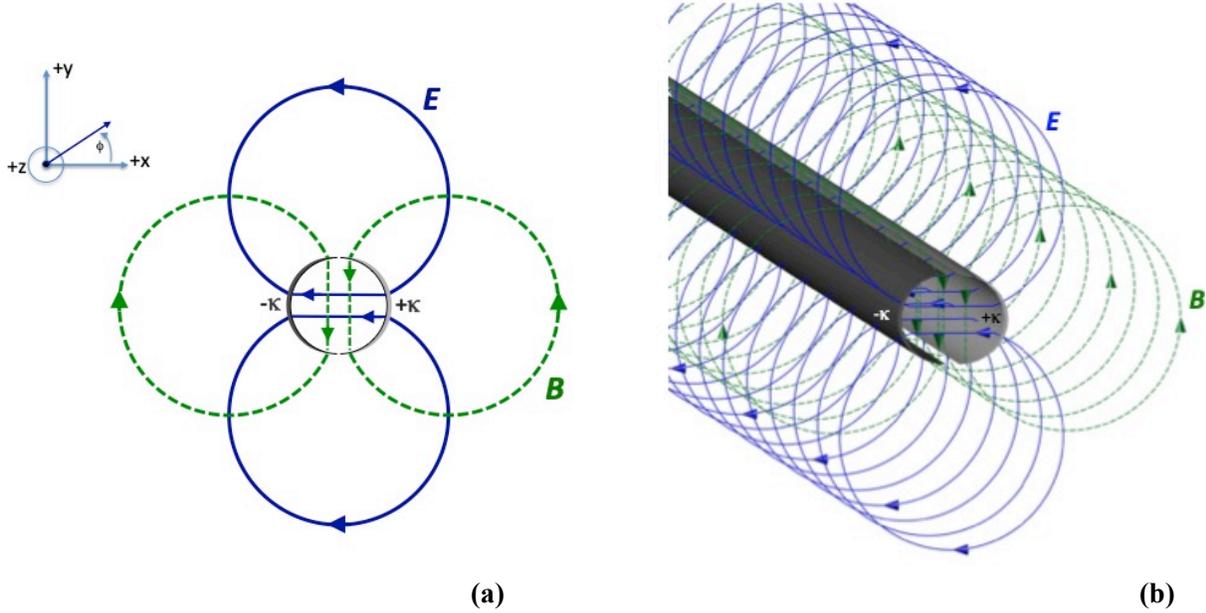

**(a)**                                            **(b)**

**Figure 1**: (a) A plot of (4) that illustrates a transverse section of the fields. The center ring represents the waveguide and the right and left semicircular arcs represent, respectively, layers of positive and negative polarization charges described by (5). Solid lines represent electric flux, and dotted lines represent magnetic flux. The interior flux lines are straight and intercept the guide wall at angle $\phi_0$, where $-\pi/2 < \phi_0 < \pi/2$. The exterior flux lines are circular, centered at $y = \csc(\phi_0)/2$, normal to the guide wall and, if continued through it, would pass through the axis. All electric flux lines stop and start on polarization charges and all magnetic flux lines surround $z$-directed effective polarization currents (not shown). (b) A plot of the same equation showing an extended length of waveguide.

## 3.1  A Thermodynamically closed system

Volume integrals of the field-flux and source-potential products yield, respectively, the total field energy and the field energy that remains attached to the source.  Each of the four integrals of (8), two field-flux and two source-potential integrals, has the value $W = \pi\varepsilon E_0^2 b^2 l$.  Field-flux integrals are evaluated over all space and include both free-standing flux and flux that remains attached to its source.  The source-potential integrals are evaluated over space occupied by the sources and thus include only flux that remains attached to its source.  Equality of the two forms show that there are no free-standing flux lines: all flux remains permanently attached to its source.  Therefore, a photon is a thermodynamically closed entity.  As illustrated in **Figure 1** all electric flux lines remain attached to source charges and all magnetic flux lines encircle source currents.

$$\frac{\varepsilon_0}{4}\int \boldsymbol{E} \bullet \boldsymbol{E}^* dV = \frac{1}{4\mu_0}\int \boldsymbol{B} \bullet \boldsymbol{B}^* dV = \frac{1}{4}\int \Phi\kappa^* dV = \frac{1}{4}\int \boldsymbol{A} \bullet \boldsymbol{J}^* dV \qquad (8)$$

By Thomson's theorem an isolated field ensemble cannot be stable [63].  Does the theorem extend to the above pulse and its accompanying charges?  Let $\mu_0$ represent the permeability of free space and assign a positive or negative sign, respectively, to repulsive or compressive pressures; the surface pressure $\Gamma_s$ follows from substituting the fields of (4) into the electromagnetic stress tensor and with circular polarization is:

$$\Gamma_S = \left(\varepsilon_0 E^2 - B^2/\mu_0\right)/4 = 0 \qquad (9)$$

There is no net pressure at any point on the interfacial surface: the pulse is stable.  Additional energy is required to change either the radius or the direction of the pulse.  As such the ensemble is a closed thermodynamic entity.

## 3.2  Speed of propagation

The underlying postulate of the special theory of relativity is that idealized speed of light $c$ is the same is all reference frames, with the corollary that every structure is subject to its laws.  However, considerable work has shown that not all light structures travel at speed $c$ [1-9].  Our model indicates that a photon can exist in the described form only if the entire edifice of charges and fields created by the waveforms of (4) and (5) propagate as a unit.  Consider waveform (5) as it propagates within the induced charge density that defines the interface.  The charges create a small but actual positive relative permittivity.  In response a portion of separation constant $k$ moves from the $z$-dependent wave equation, thereby decreasing the $z$-directed speed, to the transverse portion, thereby introducing $z$-dependence into the two-dimensional array of fields, leaving **Figure 1b** as an approximation to the actual result, and the propagation speed of the entire edifice at $u < c$.

Elapsed time and length in the direction of motion differ from values measured in a moving frame by the Lorentz contraction, $\Lambda$:

$$\Lambda = \lim_{u \to c}\left(1 - u^2/c^2\right)^{-1/2} \qquad (10)$$

Electromagnetic fields are also affected.  With primes indicating a fixed reference frame, electromagnetic fields in the stationary frame that are normal to the motional velocity in terms of those in the moving frame are:

$$\boldsymbol{E'} = \Lambda\left(\boldsymbol{E} - \boldsymbol{u} \times \boldsymbol{B}\right)$$
$$\boldsymbol{B'} = \Lambda\left(\boldsymbol{B} + \boldsymbol{u} \times \boldsymbol{E}/c^2\right)$$

(11)

Combining (11) with (4) shows the effective fields in the stationary frame produced by fields in the moving frame as it passes at speed $u$ are:

$$\boldsymbol{E}_{in} = 2\Lambda\boldsymbol{E}_0\left(\hat{\rho} - j\hat{\phi}\right)e^{-j\phi}e^{i(\omega t - kz)} \quad : c\boldsymbol{B}_{in} = j\boldsymbol{E}_{in}$$

(12)

$$\boldsymbol{E}_{ex} = -2\Lambda\boldsymbol{E}_0\frac{b^2}{\rho^2}\left(\hat{\rho} + j\hat{\phi}\right)e^{-j\phi}e^{i(\omega t - kz)} : c\boldsymbol{B}_{ex} = -j\boldsymbol{E}_{ex}$$

For the special case of the virtual waveguide, since the fields are normal to the direction of propagation the magnitudes are $2\Lambda$ greater than otherwise.

## 4. PHOTON CONSTRUCTION AND EMISSION

The competing interpretations of semi-classical and quantum theories of optics are nowhere starker than with spontaneous emission [64,65]. As previously discussed, the Manley Rowe equations, and more generally the nonlinear nature of oscillators [66], indicate electromagnetic field generation occurs within the nonlinear region of the atom. That region is spherical with a radius ~30 times greater than the nucleus itself, while the radius of the complete atom is ~ $10^4$ times greater than that of the nonlinear region; see **Table I**. To exit an atom radiation must first traverse the linear region of the atom with its plasma-like charge cloud.

### 4.1 Why atoms do not behave like antennas

The $Q$ of any radiating object is commonly defined as $Q = \omega W_{pk}/P_{av}$, and with all antenna radiation the least possible value is that of a dipole field, for which ~ $1/(ka)^3$ [67,68], with $k = 2\pi/\lambda$ and $a$ the radius of a virtual sphere just enclosing the antenna. With atoms emitting at optical wavelengths, $ka$ is $\approx 10^{-3}$ and hence $Q \approx 10^9$. For something the size of an atom radiating optical wavelengths the reactive, or standing, energy would be $\approx 10^9$ greater than the output power per cycle. However, while an antenna is a fixed system, often a metallic conductor, and is restricted in its ability to respond to an applied source, an atom is an adaptive system with no such physical constraints and responds to local force fields. Adaptive systems minimize energy and minimization of standing energy dictates the total absence of radiated power.

Consider an atom with two eigenstates between which selection rules permit energy exchanges: high-energy eigenstate one is occupied and low-energy eigenstate two is not. The high-energy electron supports frequency $\omega_1$, see (3), and the low-energy electron is capable of supporting frequency $\omega_2$. When both states, or a portion of both states, are in the nonlinear region difference frequency $\omega$ is created; the energy is either emitted or reflected back to the source electron. Emission from an atom requires the field to move into and through the linear region, radius ~100 fm to ~100 pm, with its adaptable electron charge.

With $\varepsilon$ representing permittivity, after the field enters the linear region the dipole field has the form:

$$\boldsymbol{E} = \frac{p}{4\pi\varepsilon r^3}\left\{2\sin\theta\,\hat{r} - \left(\cos\theta\,\hat{\theta} \pm j\hat{\phi}\right)\right\}e^{\pm j\phi}e^{i(\omega t - kr)}$$

(13)

The electron-cloud responds to any and all entering fields by generating an electric dipole field that is identical in all respects except magnitude and phase: the newly formed field is π out-of-phase with the incoming field and reflects the applied energy back to the source. The same process applies to any and all higher order multipole fields [69]. Supplemental Information SI-1 presents a third approach for understanding why such fields are not observed.

Moore penned a re-constructed conversation between Bohr and Schrödinger in which Schrödinger explained the type of signal he would expect during 'quantum jumps', and the signals should be large enough to be detected outside the atom, but they are not [70]. The above procedure describes the extinguishing process applicable to all wave forms that interact with the atom's complement of electrons, hence the absence of 'quantum jump' radiation.

### 4.2 Photon construction by an atom

Critical photon-creation events necessarily occur during a time period not to exceed the time for a propagating field to exit the atom, ~$10^{-19}$ s, hence a complete mathematical description includes transient solutions of the electromagnetic equations, and they are not available. Therefore, by default, our analysis is based upon a steady-state description of the optical frequency radiation that first enters linear space, and the contained pulse as it traverses the atom's linearly responding, electron-filled region.

The many mutual characteristics of circular dielectric waveguides and photons are discussed in **Section 3.1**, and lead us to closely examine if natural processes expected within atoms can create an equivalent waveguide. An important theorem of classical electromagnetics applicable to linear media is that the source of every electromagnetic field may be expressed as a sum over its own unique set of multipolar fields. At optical frequencies the $ka$ ratio of atoms, and more so the nuclear region, is so small that only the first expansion term of electric dipole radiation provides a significant output, and that requires order and degree modal numbers (1,±1) [69].

By our model, for optical radiation to exit the atom the three-dimensional expanse of the dipole field generated by transition of energy from one eigenstate to another must compress into a TEM mode propagating within a one-dimensional waveguide that protects it from outside influences as it traverses the electronic portion of the atom. Optical selection rules require the photon to carry angular momentum; this, in turn, requires rotating dipole moments $\boldsymbol{p}$ that are described by both:

$$\boldsymbol{p} = p\left(\hat{x} \mp i\hat{y}\right)e^{i\omega t} = p\left(\hat{\rho} \mp i\hat{\phi}\right)e^{i\left(\omega t \mp \phi\right)} \tag{14}$$

The circularly polarized electric dipole fields that uniquely satisfy these requirements, written with scalar value of $p$, are:

$$\boldsymbol{E} = \frac{p}{4\pi\varepsilon r^3}\left\{2\sin\theta\,\hat{r} - \left(\cos\theta\,\hat{\theta} \mp i\hat{\phi}\right)\right\}e^{i\left(\omega t - kr \mp \phi\right)} \tag{15}$$

We next consider if field (15) can generate a waveguide-like structure of charge with radius $b$ that binds and guides the fields. For that purpose, it is convenient to re-express it using a mix of spherical $(r, \theta, \phi)$ and $(\rho, \phi, z)$ cylindrical coordinates. Including the static nuclear field $E_N$, our choice for describing the total electric fields are the mixed coordinate forms:

$$\boldsymbol{E} = \frac{p}{4\pi\varepsilon r^3}\left\{3\sin\theta\,\hat{r} - \left(\hat{\rho} \mp i\hat{\phi}\right)\right\}e^{i\left(\omega t - kr \mp \phi\right)} + E_N\hat{r} \tag{16}$$

To detail the analysis, we choose an optical wavelength of 500 nm (frequency $\nu = 600$ THz, period $\tau = 1.7$ fs). Since propagation time for light to traverse the linear region is about $\sim 10^{-19}$ s, approximately $10^4$ times less than the period of the wave, for this analysis we consider the dipole field to be static. We consider (16) immediately after it is formed at radius $r$ differentially greater than $R_S$, and hence $E_N \cong E_S$. Since our only source of a charge density that could serve as a cylindrical waveguide is through the divergence of an electric field we note, after defining $E_0 = -p/4\pi\varepsilon r^3$, that the actual dipole field of (16) on and near the z-axis is:

$$\boldsymbol{E} = E_0\Big[\hat{\rho}\cos\big(\omega t - kz \mp \phi\big) + \hat{\phi}\sin\big(\omega t - kz \mp \phi\big)\Big] \qquad (17)$$

We anticipate the magnitude of $E_0$ to be less than but comparable to $E_N$, see SI-1. The total field magnitude on and near the z-axis is $(E_0^2 + E_N^2)^{1/2} > E_S$ and large enough to induce charge density:

$$\kappa = \varepsilon_0 E_0 \cos\big(\omega t - kz \mp \phi\big) \qquad (18)$$

Inspection of (17) at field points radius $b$ and angular positions $\phi$ and $\pi + \phi$ shows the field symmetry is:

$$\boldsymbol{E}_0\big(b\big) = \boldsymbol{E}_0\big(-b\big) \qquad (19)$$

The field has even parity. Next consider the static nuclear field at the same field points. Field vectors to each point from the z-axis have the symmetry:

$$\boldsymbol{E}_N\big(b\big) = -\boldsymbol{E}_N\big(-b\big) \qquad (20)$$

The field has odd parity.

At $r$ differentially greater than $R_S$, (19) shows that construction of an appropriate charge-waveguide wall requires even parity but (20) shows the nuclear field, required to obtain a total field in excess of the threshold field, has odd parity. With $\theta_0$ the angle from the origin (centered on the nucleus) to points $b$, a suitable wave-guiding charge density can be formed by satisfying the inequality:

$$|\, E_0 \cos\big(\omega t - kR_S - \phi\big)\,| >> E_N \sin\theta_0 \qquad (21)$$

Conditions for charge induction are $\sin\theta_0 < E_0/E_N$ and a wave phase angle small enough so $\cos\big(\omega t - kR_S \mp \phi\big)$ is $\approx$ one. Since $E_N \sin\theta$ vanishes on the z-axis there must be a value of $b$ for which the inequality is satisfied. Charge is induced within the disk of radius $b$, which becomes a ring of charge density proportional to $\cos\phi$ via the process described in **Section 2.1**.

Formation of the ring of charge changes the fields from three-dimensional to one-dimensional with a significant difference in boundary conditions. Matching the altered boundary conditions changes the fields to that of (22):

$$\boldsymbol{E} = E_0\big(\hat{\rho} \mp i\hat{\phi}\big)e^{i\big(\omega t - kz \mp \phi\big)} \; : c\boldsymbol{B} = i\boldsymbol{E} \qquad (22)$$

The stage is set for photon propagation.

The initial charge induction was enabled by static field $E_N$, and $E_N$ decreases by $1/r^2$ and therefore effective for only a relatively small distance from the nucleus. In its place, with a propagation speed approaching $c$, the relativistically-augmented magnitude of $E_0$ equals or exceeds $E_S$, see (12), and it forces charge induction. Since the waveguide-cylinder is lossless and hence retains all field energy, $E_0$ is constant. The fields of (22) have the exact form of (4) and satisfy the requirement of zero divergence throughout the region. At radius $b$ the induced charge density is that of (18) and the surface current density is:

$$\eta \boldsymbol{I} = \hat{z} E_0 \cos\left(\omega t - kz \mp \phi\right) \tag{23}$$

Together, (18) and (23) describe surface effects on a circular, dielectric waveguide of radius $b$ that isolates interior fields from exterior influences. For this reason, and unlike all other fields, this specific edifice of field and charge does not suffer the fate discussed in **Section 4.1** but propagates through the linear, electron-containing portion of the atom and, without incident, propagates outward into free space. The photon continues to induce charges as it propagates. It also rotates once each field cycle and, by doing so, creates alternating bands of positive and negative charge that form into a double helix.

Since the surface charge densities created by the internal and external TEM fields of (4) are identical, it is possible for the surface charge density created by the fields of (22) to also create the exterior fields of (4). Doing so photon energy is conserved, momenta are unaffected, and both the electric field intensity at the interface and the induced charge density decrease by a factor of $\sqrt{2}$.

To summarize, although subject to chaotic disturbances the field of a propagating wave must be generated by two eigenstates at least partially within the nonlinear region and endure long enough to complete the transition. After the field has been created it must retain an appropriate form until all available energy transfers into the linear region. After entering the linear region, propagation requires the initial formation of the ring of charge by fields $E_0$ and $E_N$, and the subsequent onset of the magnetic and electric fields arising from the modified boundary conditions. The magnitude of $E_0$, when appropriately modified by relativistic considerations, must be large enough so that when propagating at a speed approaching $c$ induced charges will be continuously formed. The cumulative effect of these uncertainties is that the onset of spontaneous emission is probabilistic.

As per the known absence of linearly polarized photons, linearly polarized dipole fields do not induce an enclosing, field-protecting charge structure and hence cannot propagate in accordance with the discussion of **Section 4.1**.

### 4.3 Photon size estimates

As noted in **Section 3.2** a photon propagates at a speed less than $c$; the exact value is unknown yet photon characteristics depend upon it. Photon parameters of interest are the electric field intensity $E_0$, radius $b$, $l$ the photon length in its own reference frame, and the fractional wavelength $l/\lambda$. The energy-size relationship is shown in (6), and given by $W = \pi \varepsilon_0 E_0^2 b^2 l = h\nu/2$.

Continued induction of the photon-enclosing cylindrical charge array requires that $\Lambda E_0 = E_S$; it is convenient to introduce new variable, speed ratio $\alpha$ where:

$$u/c = \left(1 - 10^{-\alpha}\right) \ : \ \Lambda \cong 10^{\alpha/2}/\sqrt{2} \tag{24}$$

The dipole field intensity is:

$$E_0 = \sqrt{2}\, E_S \times 10^{-\alpha/2} \tag{25}$$

These equations are adequate to construct **Table II**.

**Table II**: Relationships between photon parameters and size.

| $b$ (am) | $\alpha$ | $E_0$ (V/m) | $l\,/\,\lambda$ | $l$ (nm) |
|---|---|---|---|---|
| 10 | 4 | $1.84\times10^{16}$ | 0.42 | 210 |
| 100 | 2 | $1.84\times10^{17}$ | 0.42 | 0.021 |
| 10 | 2 | $1.84\times10^{17}$ | 0.0042 | 2.1 |
| 100 | 4 | $1.84\times10^{16}$ | 0.0042 | 2.1 |

Column one is photon radius $b$ in attometers, column two is speed ratio $\alpha$, column three is field intensity $E_0$ calculated from (26), column four is ratio of photon length $l$ determined in its own reference frame to wavelength for $\lambda = 500$ nm, and column five is the photon length observed as the photon passes. Nature's means of coding information about photon frequency bears on acceptable photon sizes. For example, if coding is by energy the ratio $l\,/\,\lambda = 0.0042$ may be acceptable, but if coding is by wavelength a ratio of at least $l\,/\,\lambda = 0.42$ must surely be required.

**4.4 Stimulated emission and absorption**

Without transient solutions of the electromagnetic equations we can only outline a few parameters that will surely be significant for the process. By our model of an atom the activating photon must penetrate the nonlinear region to incite an energy exchange; it is there that all energy exchanges occur. Therefore, the photon must penetrate the atomic volume and traverse inwardly through the linear region. The ratio of the linear-to-nonlinear radii, 100 pm to 100 fm, is about 1000; as an example, iron has an estimated nuclear radius of 4000 am and, by **Table II**, photon radius $b$ is in the 10 to 100 am range. It is 'needle-like' even on a nuclear scale of dimensions. The small size is surely significant and helpful in penetrating the atom. Since there is no time delay between the onsets of incoming radiation and a stimulated output, we conclude there are no time-delaying probabilistic events required to complete the exchange, such as those required for spontaneous emission, from which we conclude the incoming photon acts as a template for its daughter photon.

**5. ENTANGLEMENT**

Since two photons described by a single wave function occurs most frequently with near-neighbors, we use our photon model to calculate the transverse force between a pair of neighboring photons propagating along parallel paths. If inter-photon forces do not adjust the charge structures, we show a small attractive force is created between pairs with antiparallel spins, and a sinusoidally time-varying force is created between pairs with parallel spins.

As noted in **Section 4.2** with atomic values of $ka$ only the lowest order term in an expansion for the dipole field need be retained [69]. Our concern here is any force that may exist between near-neighbor photons propagating on parallel paths that are identical in all respects except the spins may be either parallel or antiparallel. Step by step details are given in **SI-3**.

Defining $\chi = \omega t - kz$ our concern is the outer phasor fields of (4) which implicitly travel at $c$ :

$$\boldsymbol{E} = \boldsymbol{E}_0\,\frac{b^2}{\rho^2}\left(-\hat{x}\pm i\hat{y}\right)e^{i(\chi\pm2\phi)} \quad : c\boldsymbol{B} = \boldsymbol{E}_0\,\frac{b^2}{\rho^2}\left(\mp i\hat{x}-\hat{y}\right)e^{i(\chi\pm2\phi)} \qquad (26)$$

We can calculate the force most directly by using the actual parts of the fields:

$$\boldsymbol{E} = E_0 \frac{b^2}{\rho^2} \big[ -\hat{x}\cos\left(\chi \pm 2\phi\right) \mp \hat{y}\sin\left(\chi \pm 2\phi\right) \big]$$

$$c\boldsymbol{B} = E_0 \frac{b^2}{\rho^2} \big[ \pm\hat{x}\sin\left(\chi \pm 2\phi\right) - \hat{y}\cos\left(\chi \pm 2\phi\right) \big]$$

(27)

Re-expressing (27) using trigonometric functions of single variables gives:

$$\boldsymbol{E} = E_0 \frac{b^2}{\rho^2} \big[ \hat{x}\left\langle -\cos 2\phi\cos\chi \pm \sin 2\phi\sin\chi \right\rangle - \hat{y}\left\langle \pm\cos 2\phi\sin\chi + \sin 2\phi\cos\chi \right\rangle \big]$$

$$c\boldsymbol{B} = E_0 \frac{b^2}{\rho^2} \big[ \hat{x}\left\langle \pm\cos 2\phi\sin\chi + \sin 2\phi\cos\chi \right\rangle + \hat{y}\left\langle -\cos 2\phi\cos\chi \pm \sin 2\phi\sin\chi \right\rangle \big]$$

(28)

We seek the force between two near-neighbor photons that are identical in all respects, with the two photons having either parallel or antiparallel spins. They propagate in the +z-direction and are spaced $\Delta x = 2d$ between centers. Consider a virtual planar strip $y\Delta z$ through $x = 0$ that extends between $y = \pm\infty$. The fields expressed in (28) are those of a photon at $(-d,y)$. The photon located at position $(+d,y)$ creates the fields of (29):

$$\boldsymbol{E} = E_0 \frac{b^2}{\rho^2} \big( \hat{x}\left\langle -\cos 2\phi\cos\chi \mp \sin 2\phi\sin\chi \right\rangle - \hat{y}\left\langle \pm\cos 2\phi\sin\chi - \sin 2\phi\cos\chi \right\rangle \big)$$

$$c\boldsymbol{B} = E_0 \frac{b^2}{\rho^2} \big( \hat{x}\left\langle \pm\cos 2\phi\sin\chi - \sin 2\phi\cos\chi \right\rangle + \hat{y}\left\langle -\cos 2\phi\cos\chi \mp \sin 2\phi\sin\chi \right\rangle \big)$$

(29)

For photons with parallel spins the summed fields on the strip are:

$$\boldsymbol{E}_{\uparrow\uparrow} = E_0 \frac{2b^2}{\rho^2} \big( -\hat{x}\cos 2\phi\cos\chi \mp \hat{y}\cos 2\phi\sin\chi \big)$$

$$c\boldsymbol{B}_{\uparrow\uparrow} = E_0 \frac{b^2}{\rho^2} \big( \pm\hat{x}\cos 2\phi\sin\chi - \hat{y}\cos 2\phi\cos\chi \big)$$

(30)

For photons with antiparallel spins the summed fields on the strip are:

$$\boldsymbol{E}_{\uparrow\downarrow} = -\hat{x}E_0 \frac{2b^2}{\rho^2} \big( \pm\cos 2\phi\cos\chi + \sin 2\phi\sin\chi \big)$$

$$c\boldsymbol{B}_{\uparrow\downarrow} = -\hat{y}E_0 \frac{2b^2}{\rho^2} \big( \cos 2\phi\cos\chi \pm \sin 2\phi\sin\chi \big)$$

(31)

The electromagnetic stress tensor shows the separation pressure at each point on the strip is proportional to the square of the y-directed field component minus the square of the x-directed field component. Inspection shows that, in both cases, there is no field pressure and hence no force between adjacent photons. So long as the spacing satisfies the inequality $d > b$, photons may be packed together arbitrarily closely. This is consistent with the Bose-Einstein condition that unlimited numbers of photons may be packed into a single quantum state.

However, the conclusion of no force between the photons is correct only for photons travelling at speed $c$ and, as discussed in **Sections 3.2** and **4.2**, $u < c$. For a more precise determination of the forces we replace $c$ by $u$ in field terms (30) and (31). Substituting (24) into (30) and (31) and solving for the pressure shows:

$$P_{\uparrow\uparrow} = 4\varepsilon b^4 E_0^2 \cdot 10^{-\alpha} \left( \frac{y^2 - d^2}{\rho^4} \right)^2 \cos 2\chi$$

$$\tag{32}$$

$$P_{\uparrow\downarrow} \cong -4\varepsilon b^4 E_0^2 \cdot 10^{-\alpha} \left\{ \left( \frac{y^2 - d^2}{\rho^4} \right)^2 \cos^2\chi \pm \frac{2}{\rho^4} \left( \frac{2dy}{\rho^2} \right) \left( \frac{d^2 - y^2}{\rho^2} \right) \sin\chi \cos\chi + \left( \frac{2dy}{\rho^4} \right)^2 \sin^2\chi \right\}$$

Integrating over the entire strip shows that there are small but significant forces in the two cases:

$$F_{\uparrow\uparrow} = -\pi\varepsilon \left( \Delta z \right) \frac{b^4}{d^3} E_0^2 \cdot 10^{-\alpha} \cos 2\chi$$

$$\tag{33}$$

$$F_{\uparrow\downarrow} = -\pi\varepsilon \left( \Delta z \right) \frac{b^4}{d^3} E_0^2 \cdot 10^{-\alpha}$$

Inter-photon forces differ only in that with parallel spins it oscillates at twice the photon frequency and with antiparallel spins it is constant and attractive. We note that the attractive force between antiparallel spins may play a role in entanglement [11-14]. The guiding and confining charges lie in an intermediate range between positive and negative energies, not entirely in either, and hence remain attached to Dirac's negative energy sea of charge, and there is no known speed limitations for signals between negative energy states.

## 6. CONCLUSIONS AND DISCUSSSION

The key points of this work are based upon the postulate that negative-energy charge adapt positive energy attributes when subject to static, as well as dynamic [18], electric field intensity of $1.3 \times 10^{18}$ V/m, the threshold field. Both static and dynamic threshold-level electromagnetic fields force a nonlinear response from the spatial vacuum that induces negative-energy charges to adapt positive energy characteristics and thus prevent any field from exceeding the threshold value. Jackson noted static nuclear fields should be as large as $10^{21}$ V/m [30]. Our postulate enables us to model an atom's nuclear region as a positive nucleus surrounded by induced negative charge in the form of a uniform, nucleus-adjoining layer. An equal amount of induced positive charge exists arrayed throughout the atom, concentric sphere. The field inside the sphere is equal to the threshold field and the vacuum within responds nonlinearly to applied fields.

With spontaneous emission, the filled high-energy and the empty low-energy states can only exchange energy when parts of both are in the nonlinear region in accordance with the Manley-Rowe equations [31]. Optical frequency energy is created and located in the immediate vicinity of the nucleus. Our view of photon formation and emission by the atom is discussed in **Section 4**; it requires that linear space support high intensity electromagnetic waves and that nonlinear space induces the layers of charge that guide the waves. After emission, arrays of field-induced, vacuum charge form an equivalent optical waveguide that guides and confines the energy as it travels endlessly away from the atom. All participating charge remains *in situ*: none propagates and hence there is no rest mass. During propagation fields at the pulse front, enlarged by Lorentz relativistic contraction, continuously induce *in-situ* arrays of positive-energy, charged pairs, the charges remain in position during pulse passage after which they losslessly drop back to the negative-energy state.

The charge-field photon ensemble described herein, **Section 3**, propagates at a speed approaching $c$ and possesses the following unique properties: {1} the propagating ensemble has no unbalanced charge and no rest mass, {2} energy-to-linear momentum ratio is $c$, {3} energy-to-angular momentum ratio is $\omega$. The waveguide is a $z$-directed circular cylinder of an induced charge layer proportional to $\cos\phi$. With propagation, the cylinder rotates once each field cycle by which the bands of positive and negative charge form a double helix. The photon structure permits calculation of the force between neighboring photons, separately calculated for parallel and antiparallel spins. The pair with parallel spins suffers a small sinusoidal force that increases the separation with each cycle. The pair with antiparallel spins is subject to a constant attractive force that entangles the pair unless disturbed by outside forces.

Only nonlinearities can create the deviations from equilibrium that constitute an electromagnetic particle. The photon particle is a permanent thermodynamic entity consisting of TEM waves propagating past a pair of induced +/− arrays of vacuum charge that confines and guides the propagation. The uniqueness theorem assures that only fields (4) support a photon's full assortment of electromagnetic and kinematic photon properties. The energies of potential-charge and field-flux products are equal and shown in (8) and (9); and requires that all fields possess the static property of remaining attached to their sources.

Knowledge of chaos, chaos-like activity, the Schwinger threshold field, the Dirac vacuum, and the details of constraints on radiation by electrically-small antennas were not available when Einstein wrote that only nonlinear field equations can create the deviations from equilibrium that constitute an electromagnetic particle [71]. The arrays of charged pairs induced by Schwinger's vacuum nonlinearity enable the function Einstein described when he wrote to Sommerfeld in 1909 of the 'ordering of the energy of light around discrete points that move with the velocity of light' [72]. As our work shows, the photon is comprised of fields of sufficient magnitude to induce charges that, in turn, guide and bind the fields.

With our analysis it is not quantum mechanics that underlies chaos, but chaos that underlies quantum mechanics. Free electrons pulsate at frequency $\omega = mc^2/\eta$; with trapped electrons the frequency is decreased by atomic binding energies. Within linear media the response to such cycles is repetitive and readily predictable, see SI-2, but within nonlinear media the response continually varies and is probabilistic. Experimental measurements of atomic systems yield probabilistic responses; it follows that measurements yield exact answers at the moment the measurement is taken. During the process of taking a series of measurements the atomic electrons continuously evolve, influenced by properties of both linear and nonlinear regions, and yield exact but different answers at each time slot: a summation of all such readouts is probabilistic. Our ability to describe photon creation and emission using classical physics suggests statistical analysis based upon chaotically induced motion of its parts underlies quantum mechanics. After a sufficient time and an extended number of pulses, solutions become chaotic attractors. They are also the probabilistic answers that result from Schrödinger's equation; since any two things equal to the same thing are equal to each other, it follows that solutions of Schrödinger's equation are chaotic attractors.

**Acknowledgements**

The authors gratefully acknowledge the help of Monica Claire Flores of Daegu Gyeongbuk Institute of Science and Technology (DGIST) with the figures. CAG acknowledges early-stage support of this work by the Air Force Office of Scientific Research, Contract F49620-96-1-0353.

**Data Availability**
All datasets presented in this study are included in the article/supplemental information.

**Supplemental Information**

**1. Interior atomic field intensities**

To examine effects of optical frequency oscillations within atoms, we examine fields created by an oscillating hollow, spherical, charged, conducting shell. A small oscillation distance $\delta$ creates frequency dependent dipolar surface current and charge densities. The radius is $a$, where $a \ll \lambda$, $\eta$ is the wave impedance, $k$ is the wave number, $\omega$ is the radian frequency, and it supports charge $q$. When the shell undergoes charge displacement $\delta$ the dipole moment is: $p = q\delta$. Subscripts 'ex' and 'in' represent respectively external and internal to the shell. The fields are:

$$\boldsymbol{E}_{ex} = \frac{p}{4\pi\varepsilon_0 r^3}\left[2\hat{r}\cos\theta + \hat{\theta}\sin\theta\right]e^{i(\omega t - kr)} \quad : \quad \boldsymbol{B}_{ex} = \frac{i\eta p}{4\pi r^2}k\hat{\phi}\sin\theta e^{i(\omega t - kr)}$$

$$(1.1)$$

$$\boldsymbol{E}_{in} = \frac{p}{4\pi\varepsilon_0 a^3}\left[\hat{r}\cos\theta - \hat{\theta}\sin\theta\right]e^{i\omega t} \quad : \quad \boldsymbol{B}_{in} = \frac{i\eta p}{8\pi a^3}kr\hat{\phi}\sin\theta e^{i(\omega t - kr)}$$

If an atom is driven by an applied field of relatively small magnitude, and if the nuclear attractive force remains dominant, it oscillates as a single unit. This creates equal magnitudes of positive and negative dipole moments and thus there is no external field. Generally speaking, however, displacement $\delta$ may be different in different parts of the atom, and hence the atom has a net unbalanced dipole moment and generates an external field. For sources much smaller than a wavelength the field absorbs standing energy and the partial of that energy with respect to the dipole moment is a generalized force that acts to equalize its positive and negative parts. This is easily done by adjusting the relative positions of electrons or their paths, and radiation ceases. A similar but larger force acts to eliminate all higher order modes.

To illustrate, simplify an atom into two concentric spherical shells with the positive and negative charges, respectively, on the inner and outer shells. The radius of the inner shell is 50 fm and the outer shell radius is 50 pm; each has the atomic number of charges.

The field intensities of (1.1) created by each shell at $r = 50$ fm are:

$$\boldsymbol{E}_{ex} = \frac{p}{4\pi\varepsilon_0 r^3}\left[2\hat{r}\cos\theta + \hat{\theta}\sin\theta\right]e^{i(\omega t - kr)} = 1.2 \times 10^{18}\left[2\hat{r}\cos\theta + \hat{\theta}\sin\theta\right]e^{i(\omega t - kr)} Vm^{-1}$$

$$(1.2)$$

$$\boldsymbol{E}_{in} = -\frac{p}{4\pi\varepsilon_0 a^3}\left[\hat{r}\cos\theta - \hat{\theta}\sin\theta\right]e^{i(\omega t)} = -1.2 \times 10^9\left[\hat{r}\cos\theta - \hat{\theta}\sin\theta\right]e^{i(\omega t - kr)} Vm^{-1}$$

Since the fields are additive effects of the positive charges are significant while that of the negative charges are negligible.

The field intensities of (1.2) created by each shell at $r = 50$ pm are:

$$E_{ex} = -E_{in} = \frac{p(a)}{4\pi\varepsilon_0 r^3}\left[2\hat{r}\cos\theta + \hat{\theta}\sin\theta\right]e^{i(\omega t - kr)} = 1.2\times10^9\left[2\hat{r}\cos\theta + \hat{\theta}\sin\theta\right]e^{i(\omega t - kr)}$$

$$(1.3)$$

$$E_{ex} + E_{in} = 0$$

For the special case of (1.3) the fields are additive, equal in magnitude, opposite in sign, and sum to zero.

This model shows that a nascent atom supports no external field, but it does support a dipolar field near the nucleus of a magnitude comparable to the threshold field for nonlinearity but is not detectable by external measures.

## 2. Electromagnetic fields within linear and nonlinear cavities

Inside a hollow cavity the electromagnetic fields are describable only by terms that remain finite at the origin. Such functions that are also solutions of the wave equation are spherical Bessel functions, $j(\sigma)$ and $j^\bullet(\sigma)$, where $\sigma = kr$ and:

$$j(\sigma) = \frac{1}{\sigma}\left(\cos\sigma + \frac{1}{\sigma}\sin\sigma\right) \; : \; j^\bullet(\sigma) = \frac{1}{\sigma}\left(\frac{1}{\sigma}\cos\sigma + \left\langle 1 - \frac{1}{\sigma^2}\right\rangle\sin\sigma\right)$$

The fields are:

$$\tilde{E} = \frac{pk^3}{4\pi\varepsilon_0}\left\{\frac{2}{\sigma}j(\sigma)\cos\theta\,\hat{r} - j^\bullet(\sigma)\sin\theta\,\hat{\theta}\right\}\sin\omega t \; : \; \tilde{B} = \frac{\eta pk^3}{4\pi}j(\sigma)\sin\theta\,\hat{\phi}\cos\omega t \qquad (2.1)$$

As times passes continuous, repetitive, identical field cycles continue without limit. With $a$ equal to radius of the spherical cavity, the tangential component of the electric field at $\sigma = ka$ must vanish. This requires that $j^\bullet(ka) = 0$ and the smallest radius that meets the condition is $a = 0.437\lambda$. Therefore the lowest possible eigenfrequency is:

$$a/\lambda > 0.437 \; : \; \text{If } \lambda = 500\,nm, \text{ and then } a \geq 218\,nm \qquad (2.2)$$

The conclusion is atoms are too small to store optical wavelengths.

Define a new radius $b < a$ and let the region $r < b$ be occupied by a lossless, nonlinear medium but the medium in region $b < r < a$ remains linear. The surface cavity is initially excited by (2.1). Again the tangential component of the electric field at $\sigma = ka$ vanishes, and again all energy is retained within the region.

As an example, if the nonlinearity responds as the cube of the applied field it will include field terms proportional to the cube of the trigonometric functions:

$$4\cos^3\omega t = 3\cos\omega t + 3\cos\omega t \; : \; 4\sin^3\omega t = -\sin3\omega t + 3\sin\omega t \qquad (2.3)$$

The energy associated with these terms continuously interacts within the nonlinear region, and produce an ever-widening randomized mix of frequencies. The energy output per term is unpredictable but is expected to undergo continuous changes.

Our interest is how these chaotic actions affect electrons. An electron has no known components or substructure, including no measurable radius, but does have specified values of charge, magnetic moment, mass, and angular momentum. In this work we discard the common view of electrons as a hard particle, or as a wave, but instead view an electron's charge in the form of a cloud of charge. When free of constraints surface tension forms it into a minute sphere but when imprisoned within an eigenstate it expands to occupy the entire state, in accordance with the wave function. The electrons sustain time and space average values of the kinematic variables energy, linear momentum, and angular momentum. In common with an axiom of statistical mechanics, the cloud continuously evolves through all possible space and momentum distributions that satisfy the conditions. We suggest that interactions within the nonlinear region continuously randomize events within each eigenstate's cloud of charge.

### 3.0. Second order inter-photon forces

We begin with the fields of (28) as the actual, external fields created by a photon at radius $\rho > b$. As an aid in calculating the force between two photons, we construct a virtual strip that passes through the $xy$-axis for length $\Delta z$ and extends along $x = 0$ between the limits $y = \pm\infty$. The initial photon is $z$-directed, centered at position $(-d,0,0)$, and the fields on the strip are those of (28), where the double signs describe the fields of photons with oppositely directed spins. Next place at position $(d,0,0)$ a photon that is identical to the first in all respects except it may have either equal or opposite spin. The fields at angle $\phi$ from the first photon are at angle $(\pi-\phi)$ from the second. Note that $\cos[2(\pi-\phi)] = \cos2\phi$ and $\sin[2(\pi-\phi)] = -\sin2\phi$, modify (28) accordingly, and get (29) as the field produced by the second photon. The sum of (28) plus (29) is the total field at each point on the strip, and is given by (30) and (31) for the possible combinations, and repeated here at (3.1) and (3.2):

For photons with parallel spins the summed fields on the strip are:

$$\boldsymbol{E}_{\uparrow\uparrow} = E_0 \frac{2b^2}{\rho^2}\left(-\hat{x}\cos2\phi\cos\chi \mp \hat{y}\cos2\phi\sin\chi\right)$$

$$c\boldsymbol{B}_{\uparrow\uparrow} = E_0 \frac{2b^2}{\rho^2}\left(\pm\hat{x}\cos2\phi\sin\chi - \hat{y}\cos2\phi\cos\chi\right)$$

(3.1)

For photons with antiparallel spins the summed fields on the strip are:

$$\boldsymbol{E}_{\uparrow\downarrow} = -\hat{x}E_0 \frac{2b^2}{\rho^2}\left(\pm\cos2\phi\cos\chi + \sin2\phi\sin\chi\right)$$

$$c\boldsymbol{B}_{\uparrow\downarrow} = -\hat{y}E_0 \frac{2b^2}{\rho^2}\left(\cos2\phi\cos\chi \pm \sin2\phi\sin\chi\right)$$

(3.2)

The fields (28) and (29) lead to those of (3.1) by direct addition of the fields, and to the fields of (31) adding the lower double sign of (28) to the upper double signs of (29), and *vice versa*.

The electromagnetic stress tensor shows the $x$- and y-directed field components create respectively uniting and separating pressures: $P = \varepsilon(-E_x^2 + E_y^2)$. Inspection of (3.1) and (3.2) shows there is no pressure between the two photons, but this conclusion is based upon the photon speeds of exactly $c$: the conclusion of no force between the photons is correct only for photons travelling at speed $c$. As discussed in **Sections 3., 3.2**, and **4.2**, the actual speed, $u$, satisfies the inequality $u < c$. We describe actual speed in terms of empirical constant $\alpha$ of (24): $\alpha$ equals the number of nines in the expression $u = 0.99\ldots9c$, and the electric-to-magnetic field ratio of (3,1) and (3.2) is $1/(\alpha c)$. Substituting $u$ and the equalities $\rho^2 \cos 2\phi = d^2 - y^2$ and $\rho^2 \sin 2\phi = 2dy$ into the field expressions, after omitting odd functions of $y$, leaves:

$$P_{\uparrow\uparrow} = 4\varepsilon b^4 E_0^2 \bullet 10^{-\alpha} \frac{1}{\rho^8} \left( y^2 - d^2 \right)^2 \cos 2\chi$$

$$P_{\uparrow\downarrow} = 4b^4 E_0^2 \bullet 10^{-\alpha} \frac{1}{\rho^8} \left( \left( y^2 - d^2 \right)^2 \cos^2 \chi + 4d^2 y^2 \sin^2 \chi \right)$$

$$(3.3)$$

Useful integrals are:

$$\int_{-\infty}^{\infty} \frac{y^2 dy}{\rho^8} = \frac{\pi}{16d^5} \; : \; \int_{-\infty}^{\infty} \frac{dy}{\rho^4} = \frac{\pi}{2d^3} \tag{3.4}$$

Integrating the pressures over the entire strip to obtain the force between the sources gives:

$$F_{\uparrow\uparrow} = -\frac{\pi \varepsilon b^4}{d^3} \left( \Delta z \right) E_0^2 \bullet 10^{-\alpha} \cos 2\chi$$

$$F_{\uparrow\downarrow} = -\frac{\pi \varepsilon b^4}{d^3} \left( \Delta z \right) E_0^2 \bullet 10^{-\alpha}$$

$$(3.5)$$